# Thermostatistics in deformed space with maximal length

Salaheddine Bensalem[*] and Djamil Bouaziz

*LMEPA, Département de Physique, Faculté des Sciences Exactes et Informatique, Université de Jijel, BP 98, Ouled Aissa, 18000 Jijel, Algeria*

**Abstract**

The method for calculating the canonical partition function with deformed Heisenberg algebra, developed by Fityo (Fityo, 2008), is adapted to the modified commutation relations including a maximum length, proposed recently in 1D by Perivolaropoulos (Perivolaropoulos, 2017). Firstly, the formalism of 1D maximum length deformed algebra is extended to arbitrary dimensions. Then, by employing the adapted semiclassical approach, the thermostatistics of an ideal gas and a system of harmonic oscillators (HOs) is investigated. For the ideal gas, the results generalize those obtained recently by us in 1D (Bensalem and Bouaziz, 2019), and show a complete agreement between the semiclassical and quantum approaches. In particular, a stiffer real-like equation of state for the ideal gas is established in 3D; it is consistent with the formal one, which we presented in the aforementioned paper. By analyzing some experimental data, we argue that the maximal length might be viewed as a macroscopic scale associated with the system under study. Finally, the thermostatistics of a system of HOs compared to that of an ideal gas reveals that the effects of the maximal length depend on the studied system. On the other hand, it is observed that the maximal length effect on some thermodynamic functions of the HOs is analogous to that of the minimal length, studied previously in the literature.

[*]Corresponding author: bensalems@yandex.com (Salaheddine Bensalem)

## 1. Introduction

We recently studied the thermostatistics of an ideal gas [1] in one-dimensional quantum mechanics with a Generalized Uncertainty Principle (GUP), which implies the existence of a maximal length [2]. This study showed that, beside the similarity between the maximal and minimal lengths corrections on some thermodynamic quantities, the maximal length induces different effects and leads to a new thermodynamics. Especially, a real-like equation of state emerges naturally in this new formalism of quantum mechanics.

In this paper, we generalize our aforementioned work to three dimensions, and furthermore, we consider another important statistical system, namely, an ensemble of harmonic oscillators (HOs). Such a study is of interest since it gives further insights on the implications of the maximal length assumption, and it allows to perform a complete comparative study between the effects of the maximal and minimal lengths, by following the work of Ref. [1].

The idea of generalizing the Heisenberg Uncertainty Principle (HUP) is suggested by numerous theories in contemporary theoretical physics. This strategy is adopted to account for the existence of limit values for physical observables. For instance, a minimum length appears in quantum gravity [3], a maximum momentum is predicted in doubly special relativity [4], and a maximum length is hypothesized in cosmology due to the presence of the particle horizon or from possible nontrivial cosmic topology [2,5]. Thus, several models of GUP have been proposed in the literature.

The GUP incorporating a minimal length has been well discussed by Kempf and coworkers in their seminal work [6]. In connection with this deformed version of quantum mechanics, a lot of topics have been addressed in the literature, see for instance Refs. [7–19], and Refs. [20–22], devoted to statistical physics.

On the other hand, other GUPs have been suggested to include both a minimal length and a maximal momentum [23–28]. In this framework, numerous works have been performed to investigate the implications of such GUPs in statistical mechanics [29–31] and other physical problems [23–25,32–43].

Recently, Perivolaropoulos [2] proposed a new form of GUP in one dimension, which includes a maximum length, and studied numerically its implications on the harmonic oscillator in connection with one-dimensional Schrödinger equation.

In this work, we extend the formalism of Ref. [2] to arbitrary dimensions; then, by using the semiclassical approach of Ref. [44], we investigate in 3D statistical mechanics an ideal gas and a system of HOs. The effects of this new GUP induced for the two systems are examined and compared with those of the minimal length GUP, studied extensively in the literature.

In Sec. 2, the GUP with a maximal length is presented in 1D, and then a generalization to the $D$-dimensional case is performed. Sec. 3 introduces the semiclassical method used in deformed statistical mechanics to calculate the canonical partition function. In Sec. 4, this method is applied to the ideal gas; the thermodynamics of the system is studied and a real-like equation of state is established. In Sec. 5, the thermodynamics of an ensemble of $N$ HOs is investigated; the results are deeply analyzed by comparing them with those of the ideal gas and with the results obtained within the minimal length GUP. The paper is fenced with a conclusion summarizing the main obtained results and some prospects.

## 2. GUP with maximum length

### 2.1. One-dimensional case

The one-dimensional space maximal length GUP proposed by Perivolaropoulos has the form [2]

$$\Delta X \Delta P \geq \frac{\hbar}{2} \frac{1}{1 - \alpha \Delta X^2}, \tag{1}$$

where $\alpha = l_{\max}^{-2} = \alpha_0 \left(\frac{H_0}{c}\right)^2$ with $H_0$ is the Hubble constant, $c$ is the speed of light and $\alpha_0$ is a dimensionless parameter. This GUP implies the existence of a maximum position uncertainty

$$\Delta X_{\max} = l_{\max} = \frac{1}{\sqrt{\alpha}}, \tag{2}$$

and a minimum momentum uncertainty

$$\Delta P_{min} = \frac{3\sqrt{3}}{4}\hbar\sqrt{\alpha}. \tag{3}$$

The GUP (1) originates from the modified commutation relation

$$\left[\hat{X},\hat{P}\right] = \frac{i\hbar}{1-\alpha\hat{X}^2}. \tag{4}$$

A representation satisfying the deformed algebra (4) is also proposed in position space [2]

$$\hat{X} = x, \qquad \hat{P} = -\frac{i\hbar}{1-\alpha x^2}\frac{d}{dx}, \tag{5}$$

The scalar product is defined within this formalism as

$$\langle\psi|\varphi\rangle = \int_{-l_{max}}^{l_{max}} (1-\alpha x^2)\psi^*(x)\varphi(x)dx. \tag{6}$$

The term $(1-\alpha x^2)$ is added in the scalar product definition in order to retain the symmetry of the momentum operator [1,2].

### *2.2. Generalization to D-dimensional space*

The generalization of the one-dimensional commutation relation (4) that preserves the rotational symmetry is

$$\left[\hat{X}_i,\hat{P}_j\right] = \frac{i\hbar\delta_{ij}}{1-\alpha\hat{\mathbf{X}}^2}, \tag{7}$$

where $\hat{\mathbf{X}}^2 = \sum_{i=1}^{D}\hat{X}_i\hat{X}_i$. This algebra implies a maximal observable length and a nonzero minimal momentum uncertainty in each position coordinate. Assuming a commutative geometry, i.e.,

$$\left[\hat{X}_i,\hat{X}_j\right] = 0, \tag{8}$$

then, the Jacobi identity determines the commutation relations between the components of the momentum operator as

$$\left[\hat{P}_i,\hat{P}_j\right]=\frac{2i\hbar\alpha}{\left(1-\alpha\hat{\mathbf{X}}^2\right)^2}\left(\hat{X}_j\hat{P}_i-\hat{X}_i\hat{P}_j\right). \tag{9}$$

Within this deformed algebra, the position and momentum operators in coordinate space can be written as

$$\hat{X}_i=x_i,\qquad \hat{P}_i=-\frac{i\hbar}{1-\alpha\mathbf{x}^2}\frac{\partial}{\partial x_i}, \tag{10}$$

where $\mathbf{x}^2=\sum_{i=1}^{D}x_i x_i$.

Finally, the scalar product is generalized as

$$\langle\psi|\varphi\rangle=\int_{-1/\sqrt{\alpha}}^{1/\sqrt{\alpha}}\left(1-\alpha\mathbf{x}^2\right)\psi^*(\mathbf{x})\varphi(\mathbf{x})(dx). \tag{11}$$

where $(dx)$ stands for $dx_1 dx_2 \cdots dx_D$.

In the next section, we review the semiclassical approach of Ref. [44], and show how to calculate the canonical partition function for a given statistical system in the framework of GUP formalism.

## 3. Thermostatistics with GUP: Semiclassical approach

Let us consider a system of $N$ identical non-interacting particles in the external field $U$. The behavior of each particle may be described by the corresponding deformed Hamiltonian $H=\frac{\mathbf{P}^2}{2m}+U(\mathbf{X})$. To deal with such statistical system in the GUP scenario, two approaches may be employed, as in the ordinary (undeformed) case, namely, the quantum approach and the semiclassical one.

The quantum method is based on the following definition of the deformed single-particle canonical partition function [1,20,44]:

$$Q_1=\sum_n \exp(-\beta E_n), \tag{12}$$

where $E_n$ are eigenvalues of the deformed Hamiltonian, obtained by solving the corresponding deformed stationary wave equation, and $\beta = (k_B T)^{-1}$, with $k_B$ is Boltzmann's constant and $T$ is the temperature.

The semiclassical approach is based on the following deformed single-particle canonical partition function [44]:

$$Q_1 = \frac{1}{h^D} \int \frac{(dX)(dP)}{J} \exp(-\beta H), \tag{13}$$

where $(dX)$ and $(dP)$ stand for $dX_1 dX_2 \cdots dX_D$ and $dP_1 dP_2 \cdots dP_D$, respectively, $H$ is the classical Hamiltonian of a single-particle constituting the system, and $J$ is the Jacobian of the transformation, which links the variables $(X_i, P_i)$ and the canonically conjugated ones $(x_i, p_i)$ satisfying the Poisson brackets $\{x_i, p_j\} = \delta_{ij}$, $\{x_i, x_j\} = \{p_i, p_j\} = 0$.

The Jacobian can be read off from the deformed Poisson brackets, which are obtained from the quantum commutation relations as

$$\frac{1}{i\hbar}\left[\hat{A}, \hat{B}\right] \to \{A, B\}. \tag{14}$$

The Jacobian of the transformation in $D$-dimensions is [44] $J = \dfrac{\partial(X_1, P_1, \ldots, X_D, P_D)}{\partial(x_1, p_1, \ldots, x_D, p_D)}$.

In 1D, the Jacobian is given by [44]

$$J = \frac{\partial(X, P)}{\partial(x, p)} = \{X, P\}. \tag{15}$$

For the deformed algebra (4) corresponding to the GUP with a maximal length, the Jacobian (15) reads

$$J = \frac{1}{1 - \alpha X^2}. \tag{16}$$

Then, for a single-particle Hamiltonian

$$H = \frac{P^2}{2m} + U(X), \qquad (17)$$

the partition function (13), in 1D, takes the form

$$Q_1 = \frac{1}{h}\int (1-\alpha X^2)\exp(-\beta U(X))dX \int \exp\left(-\beta \frac{P^2}{2m}\right)dP. \qquad (18)$$

In 3D, one has the following from for the Jacobian [44]

$$\begin{aligned}J &= \frac{\partial(X_1, P_1, X_2, P_2, X_3, P_3)}{\partial(x_1, p_1, x_2, p_2, x_3, p_3)} \\ &= \{X_1, P_1\}\{X_2, P_2\}\{X_3, P_3\} \\ &\quad -\{X_1, P_3\}\{P_1, P_2\}\{X_2, X_3\} - \{X_1, P_2\}\{X_2, P_1\}\{X_3, P_3\} \\ &\quad -\{X_1, P_3\}\{X_2, P_2\}\{X_3, P_1\} - \{X_1, P_1\}\{X_2, P_3\}\{X_3, P_2\} \\ &\quad +\{X_1, X_2\}\{P_1, P_3\}\{X_3, P_2\} + \{X_1, P_3\}\{X_2, P_1\}\{X_3, P_2\} \\ &\quad -\{X_1, X_2\}\{P_2, P_3\}\{X_3, P_1\} + \{X_1, P_2\}\{X_2, X_3\}\{P_1, P_3\} \\ &\quad -\{X_1, X_3\}\{P_1, P_3\}\{X_2, P_2\} + \{X_1, X_3\}\{X_2, P_1\}\{P_2, P_3\} \\ &\quad +\{X_1, X_3\}\{P_1, P_2\}\{X_2, P_3\} - \{X_1, X_2\}\{P_1, P_2\}\{X_3, P_3\} \\ &\quad -\{X_1, P_1\}\{X_2, X_3\}\{P_2, P_3\} + \{X_1, P_2\}\{X_2, P_3\}\{X_3, P_1\}. \end{aligned} \qquad (19)$$

For the deformed algebra (7)-(9), the Jacobian (19) leads to the expression

$$J = \frac{1}{(1-\alpha \mathbf{X}^2)^3}. \qquad (20)$$

Then, for a single-particle 3D Hamiltonian

$$H = \frac{\mathbf{P}^2}{2m} + U(\mathbf{X}), \qquad (21)$$

the single-particle partition function (13) can be written in 3D as

$$Q_1 = \frac{1}{h^3}\int (dX)(1-\alpha \mathbf{X}^2)^3 \exp(-\beta U(\mathbf{X}))\int (dP)\exp\left(-\beta \frac{\mathbf{P}^2}{2m}\right), \qquad (22)$$

where $\mathbf{X}^2 = \sum_{i=1}^{3} X_i X_i$ and $\mathbf{P}^2 = \sum_{i=1}^{3} P_i P_i$.

In what follows, formulas (18) and (22) will be applied to study an ideal gas and a system of $N$ HOs.

**4. Ideal gas with a maximal length**

In Ref. [1], a detailed study of the 1D ideal gas has been performed in the context of GUP (1) by using the quantum approach. In this section, we study this system by using formulas (18) and (22) corresponding to 1D and 3D cases, respectively.

*4.1. Partition function*

Let us consider an ideal gas in 1D box of length $L$. According to (18), the single-particle partition function is

$$Q_1 = \frac{1}{h}\int_0^L (1-\alpha X^2)dX \int_{-\infty}^{+\infty} \exp\left(-\beta\frac{P^2}{2m}\right)dP, \tag{23}$$

which yields

$$Q_1 = q_1\left(1 - \frac{\alpha}{3}L^2\right), \tag{24}$$

where $q_1 = \frac{L}{\hbar}\left(\frac{m}{2\pi\beta}\right)^{1/2}$ represents the ordinary one-particle partition function. For $N$ indistinguishable particles constituting the system, one has

$$Q = \frac{Q_1^N}{N!} = \frac{q_1^N}{N!}\left(1 - \frac{\alpha}{3}L^2\right)^N = q\left(1 - \frac{\alpha}{3}L^2\right)^N, \tag{25}$$

where $q = \frac{q_1^N}{N!}$ represents the ordinary partition function of the system.

As in ordinary statistical physics, the classical partition function (25) is identical to the quantum one, derived in Ref. [1]. Such accordance between the results of classical and quantum treatments of an ideal gas has also been concluded in the framework of the GUP with a minimal length [21].

From (25), up to first order of $\alpha$, the generalized partition function can be written as

$$Q \approx q\left(1 - \frac{1}{3}\alpha N L^2\right). \tag{26}$$

Now, let us consider an ideal gas in 3D box with a volume $V = L^3$. Based on the definition (22), the partition function can be written as

$$Q_1 = \frac{1}{h^3} \int_0^L (dX)\left(1 - \alpha \mathbf{X}^2\right)^3 \int_{-\infty}^{+\infty} (dP) \exp\left(-\beta \frac{\mathbf{P}^2}{2m}\right), \tag{27}$$

the integration over the six variables of the phase space results in

$$Q_1 = q_1 F(\alpha, V), \tag{28}$$

where $q_1 = \frac{V}{\hbar^3}\left(\frac{m}{2\pi\beta}\right)^{3/2}$ is the ordinary partition function for a single particle, and $F(\alpha, V)$ denotes the deformation term given by

$$F(\alpha, V) = 1 - 3\alpha V^{2/3} + \frac{19}{5}\alpha^2 V^{4/3} - \frac{583}{315}\alpha^3 V^2. \tag{29}$$

The partition function of $N$ indistinguishable particles constituting the system, is then

$$Q = \frac{q_1^N}{N!}\left(F(\alpha, V)\right)^N = q\left(F(\alpha, V)\right)^N. \tag{30}$$

where $q = \frac{q_1^N}{N!}$ represents the ordinary partition function of an ideal gas. Up to first order of $\alpha$, one gets

$$Q \approx q\left(1 - 3\alpha N V^{2/3}\right). \tag{31}$$

Given that both partition functions (26) and (31) are similar, it is then expected to obtain the same modified thermodynamics in 1D and 3D cases, apart from a numerical factor characterizing the dimension of space.

### *4.2. Thermodynamic properties*

In Ref. [1], the thermodynamics of 1D ideal gas has been investigated in detail. Here, we consider the 3D case, and show that the results obtained in 1D case are consistent with those of the 3D case.

Let us start with the internal energy and heat capacity at constant volume

$$E = -\left(\frac{\partial \ln Q}{\partial \beta}\right)_{N,V} = \frac{3}{2} N k_B T, \quad (32)$$

$$C_V = \left(\frac{\partial E}{\partial T}\right)_{N,V} = \frac{3}{2} N k_B. \quad (33)$$

These results are similar to those obtained in 1D case [1], beside a numerical factor characterizing the dimension of space. It is to recall that, in contrast to the minimal length GUPs [23–25,27,28], the maximal length GUP does not influence the internal energy and heat capacity at constant volume of an ideal gas.

On the deformed Helmholtz free energy, it is given by

$$A = -k_B T \ln Q = a - N k_B T \ln F(\alpha, V), \quad (34)$$

here $a = -k_B T \ln q$ is the ordinary Helmholtz free energy, and $F$ is given by Eq. (29).

From Eq. (34), one can derive the entropy $S$, the chemical potential $M$, and the pressure of the system.

For the generalized entropy, one has

$$S = -\left(\frac{\partial A}{\partial T}\right)_{N,V} = s + N k_B \ln F, \quad (35)$$

where $s = -\left(\frac{\partial a}{\partial T}\right)_{N,V}$ is the ordinary entropy. From (29), up to first order of the deformation parameter $\alpha$, one obtains

$$S \approx s - 3\alpha V^{2/3} N k_B. \tag{36}$$

The generalized chemical potential is

$$M = \left(\frac{\partial A}{\partial N}\right)_{V,T} = \mu - k_B T \ln F, \tag{37}$$

where $\mu = \left(\frac{\partial a}{\partial N}\right)_{V,T}$ is the ordinary chemical potential. In the first order of $\alpha$, one has

$$M \approx \mu + 3\alpha V^{2/3} k_B T. \tag{38}$$

The corrections induced by the maximal length on these thermodynamic functions are similar to those computed in 1D case [1], beside a numerical factor, which characterizes the dimension of space. Therefore, this study confirms the validity of the conclusions reached in Ref. [1].

Let us now establish the equation of state for an ideal gas in the presence of a maximal length. Note that, a formal equation of state using 1D model has been obtained in Ref. [1].

From the definition of the pressure

$$P = -\left(\frac{\partial A}{\partial V}\right)_{N,T}, \tag{39}$$

one gets, up to $O(\alpha)$, the following equation of state:

$$PV = N k_B T \left(1 - 2\alpha V^{2/3}\right). \tag{40}$$

Again, beside a numerical factor characterizing the spatial dimension, Eq. (40) is similar to the 1D formal equation of state obtained in Ref. [1]. Therein, the features of this equation have been discussed in detail and showed that it is consistent with the real gas behavior.

### 4.3. Mayer's relation

For an ideal gas, the well known Mayer's relation reads

$$c_P - c_V = N k_B. \tag{41}$$

where, $c_P = \frac{5}{2}Nk_B$ and $c_V = \frac{3}{2}Nk_B$ are the heat capacity at constant pressure and the one at constant volume in the undeformed case, respectively.

Let us examine this formula in the presence of a maximal length. From Eqs. (32) and (40), up to $O(\alpha)$, one finds

$$C_P = \left(\frac{\partial(E+PV)}{\partial T}\right)_{N,P} = \frac{5}{2}Nk_B\left(1 - \frac{20}{15}\alpha V^{2/3}\right), \tag{42}$$

then

$$C_P - C_V = Nk_B\left(1 - \frac{10}{3}\alpha V^{2/3}\right). \tag{43}$$

This might be viewed as a generalized Mayer's relation including the maximal length effect. Formula (43) is consistent with the nonideality of gases, which manifests by the deviation of the ratio $(C_P - C_V)/N$ from its ideal value of $k_B$ [45].

To end this section, it is significant to estimate, even roughly, a lower bound for the maximal length by considering its correction on the heat capacity at constant pressure. For this purpose, we fellow the Ref. [46]. From Eq. (42), one gets

$$\frac{\Delta C_P}{c_P} = \frac{C_P - c_P}{c_P} = -\frac{20}{15}\alpha V^{2/3} \approx -\alpha_0 10^{-52} V^{2/3}. \tag{44}$$

Recall that the dimensionless deformation parameter $\alpha_0$ is related to the maximal length by $\alpha = l_{max}^{-2} = \alpha_0\left(\frac{H_0}{c}\right)^2$, where $H_0$ is the Hubble constant and $c$ is the speed of light.

Nowadays, accuracy on the experimental values of heat capacities is about $10^{-7}$ [46,47]. For a monatomic gas confined in a volume of 1 m$^3$, within this precision and by using Eq. (44), the following upper bound can be set for the parameter $\alpha_0$:

$$\alpha_0 < 10^{45}. \tag{45}$$

This upper bound is comparable to the minimal length dimensionless parameter, estimated in Ref. [46].

The lower bound for the maximal length that arises from (45) is then

$$l_{max} > 10^{7/2} \, m, \qquad (46)$$

It follows that the maximal length is size-dependent and hence, it would not be always close to the maximum measurable length in the Universe ($10^{26}$ m [2]), but it might be associated to a macroscopic scale of the system under study.

**5. Harmonic oscillators in maximal length formalism**

The system of HOs has been extensively studied in the context of the GUP with minimal length; see, for instance, Refs. [20,44]. Here, we consider this system within this new GUP incorporating a maximal length. The objective of this study is twofold: first, it allows to examine the maximal length effects on different statistical systems, and second, it completes the comparative study between the implications of the minimal and maximal lengths, started in Ref. [1], by treating a system of HOs.

***5.1. Partition function***

Let us begin by a 1D harmonic oscillator for which the Hamiltonian is given by

$$H = \frac{P^2}{2m} + \frac{m\omega^2}{2} X^2. \qquad (47)$$

From the definition (18), the single-particle partition function reads

$$Q_1 = \frac{1}{h} \int_{-l_{max}}^{+l_{max}} (1-\alpha X^2) \exp\left(-\frac{\beta m \omega^2}{2} X^2\right) dX \int_{-\infty}^{+\infty} \exp\left(-\frac{\beta}{2m} P^2\right) dP, \qquad (48)$$

where the spatial integration is performed by taking into account the existence of a maximum length. One obtains the expression

$$Q_1 = q_1 f(T, \alpha), \qquad (49)$$

with $q_1 = \dfrac{1}{\beta \hbar \omega} = \dfrac{k_B T}{\hbar \omega}$ is the ordinary partition function, and $f$ stands for the deformation term induced by the maximum length; it is given by

$$f(T,\alpha) \equiv f(\beta,\alpha) = \left(1 - \frac{\alpha}{\beta m\omega^2}\right)\mathrm{erf}\left(\sqrt{\frac{\beta m\omega^2}{2\alpha}}\right) + \frac{1}{\sqrt{\pi}}\sqrt{\frac{2\alpha}{\beta m\omega^2}}\exp\left(-\frac{\beta m\omega^2}{2\alpha}\right). \tag{50}$$

Assuming that the parameter $\alpha$ is too small such as, $\frac{\alpha}{\beta m\omega^2} \ll 1$, one has in the first order of $\alpha$

$$f(\beta,\alpha) \approx 1 - \frac{\alpha}{\beta m\omega^2}. \tag{51}$$

The partition function for $N$ distinguishable HOs takes the form

$$Q = Q_1^N = q(f(T,\alpha))^N, \tag{52}$$

where $q = q_1^N$ is the ordinary partition function of the system in 1D. It is to mention that the expression of the partition function (49) with the approximation (51) is similar to that obtained within the minimal length GUP in Ref. [44] in the "low temperature regime".

Now, let us show that the partition function of three-dimensional HOs can also be computed in the same manner. Hence, by considering the 3D Hamiltonian

$$H = \frac{\mathbf{P}^2}{2m} + \frac{m\omega^2}{2}\mathbf{X}^2, \tag{53}$$

and by using the definition (22), the partition function for one oscillator reads

$$Q_1 = \frac{1}{h^3}\int_{-l_{\max}}^{+l_{\max}}(d\mathbf{X})(1-\alpha\mathbf{X}^2)^3 \exp\left(-\frac{\beta m\omega^2}{2}\mathbf{X}^2\right)\int_{-\infty}^{+\infty}(d\mathbf{P})\exp\left(-\frac{\beta}{2m}\mathbf{P}^2\right), \tag{54}$$

the integration yields

$$Q_1 = q_1 F(T,\alpha), \tag{55}$$

where $q_1 = \frac{1}{(\beta\hbar\omega)^3} = \left(\frac{k_B T}{\hbar\omega}\right)^3$ is the ordinary partition function, and $F$ is given by

$$F(T,\alpha) \equiv F(\eta,\alpha) = \left(1 - \frac{9}{2}\frac{\alpha}{\eta} + \frac{45}{4}\left(\frac{\alpha}{\eta}\right)^2 - \frac{105}{8}\left(\frac{\alpha}{\eta}\right)^3\right)\mathrm{erf}\left[\sqrt{\frac{\eta}{\alpha}}\right]^3$$
$$+ \frac{3}{\sqrt{\pi}}\left(\sqrt{\frac{\alpha}{\eta}} - 5\left(\frac{\alpha}{\eta}\right)^{3/2} + \frac{57}{4}\left(\frac{\alpha}{\eta}\right)^{5/2}\right)e^{-\frac{\eta}{\alpha}}\mathrm{erf}\left[\sqrt{\frac{\eta}{\alpha}}\right]^2 \quad (56)$$
$$- \frac{36}{\pi}\left(\frac{\alpha}{\eta}\right)^2 e^{-2\frac{\eta}{\alpha}}\mathrm{erf}\left[\sqrt{\frac{\eta}{\alpha}}\right] + \frac{6}{\pi^{3/2}}\left(\frac{\alpha}{\eta}\right)^{3/2} e^{-3\frac{\eta}{\alpha}},$$

with $\eta = \frac{\beta m \omega^2}{2} = \frac{m\omega^2}{2k_B T}$.

Therefore, the partition function of *N* distinguishable three-dimensional HOs reads

$$Q = q\left(F(T,\alpha)\right)^N, \quad (57)$$

where $q = q_1^N$ is the 3D ordinary partition function of the system.

### 5.2. Thermodynamic quantities

As illustrated in Sec. 4, the effect of the maximal length does not depend on the spatial dimension. Thus, for the sake of simplicity, we consider in what follows the 1D partition function (52) to examine the thermodynamic properties of the system.

Let us start with the internal energy *E*, and the heat capacity at constant volume $C_V$ (V≡L). By using Eq. (52) together with Eq. (51), one can get in the first order of the deformation parameter *α* the following expressions:

$$E = -\left(\frac{\partial \ln Q}{\partial \beta}\right)_{N,V} \approx Nk_B T - \frac{\alpha N}{m\omega^2}(k_B T)^2, \quad (58)$$

$$C_V = \left(\frac{\partial E}{\partial T}\right)_{N,V} \approx Nk_B - \frac{2\alpha N}{m\omega^2}k_B^2 T. \quad (59)$$

In contrast with the ideal gas, where the maximal length GUP does not influence the internal energy and heat capacity at constant volume, in the case of HOs these quantities are affected. Moreover, Eqs. (58) and (59) indicate that, for a given *m* and *ω*, the effect of the maximal length grows at high temperatures, leading to a zero heat capacity at $T = \frac{m\omega^2}{2\alpha k_B}$. These outcomes are similar to those obtained in the presence of a minimal length; see Eqs. (44) and (45) of Ref. [20].

The generalized Helmholtz free energy can also be derived from Eqs. (52) and (51); one has in the first order of α

$$A = -k_B T \ln Q \approx Nk_B T \ln\left(\frac{\hbar\omega}{k_B T}\right) + \alpha \frac{N(k_B T)^2}{m\omega^2}, \tag{60}$$

From this equation, the entropy $S$, and the chemical potential $M$ follow straightforwardly

$$S = -\left(\frac{\partial A}{\partial T}\right)_{N,V} \approx Nk_B \left(1 + \ln\left(\frac{k_B T}{\hbar\omega}\right)\right) - 2\alpha \frac{Nk_B^2 T}{m\omega^2}, \tag{61}$$

and

$$M = \left(\frac{\partial A}{\partial N}\right)_{V,T} \approx k_B T \ln\left(\frac{\hbar\omega}{k_B T}\right) + \alpha \frac{(k_B T)^2}{m\omega^2}, \tag{62}$$

From Eqs. (61) and (62) one observes that the presence of a maximal length induces a negative correction to the entropy, and a positive correction to the chemical potential of the system, similarly to the ideal gas (see Eqs. (36) and (38)).

**6. Conclusions**

In this paper, the thermostatistics of an ideal gas and a system of HOs has been studied in a deformed space with a maximal length. Firstly, the 1D deformed algebra of Ref. [2] has been extended to arbitrary dimensions. Then, the semiclassical approach of Ref. [44], which allows for computing the canonical partition function with deformed commutation relations, has been adapted to the maximal length formalism. We applied this method to calculate the generalized canonical partition functions for both considered systems, in 1D and 3D cases.

For the ideal gas, the work of Ref. [1] has been extended to 3D; it is shown that, as expected, the semiclassical and quantum approaches provide the same results. Furthermore, a stiffer real-like equation of state for the ideal gas has been established. In addition, a generalized Mayer's relation incorporating a maximal length has been obtained. Using experimental data, a lower bound of the maximal length is estimated of about $10^{7/2}$ m, which is so far from the limit of maximum measurable length scale in the Universe ($10^{26}$ m [2]). This result leads us to guess that the maximal length would depend on the size of the system under study, and hence, it might be viewed as a scale associated to the concerned system.

For the HOs, the effect of the maximal length has been examined by probing its corrections on several thermodynamic functions. A comparison with the maximal length corrections on the ideal gas showed that the maximal length effect depends on the studied system. Furthermore, a comparison with the minimal length thermodynamics of the HOs indicates some qualitative similarity between the maximal and minimal lengths effects; which is in contrast with the conclusion reached in the case of an ideal gas, where the maximal and minimal lengths induce different effects. It would be then of interest to reexamine this issue by studying a system of quantum HOs in the presence of a maximal length. This topic will be addressed in follow-up works.

**Acknowledgments**

The work of DB is supported by the Algerian Ministry of Higher Education and Scientific Research, under the PRFU Project No. D01720140007.